\documentclass[pre]{revtex4}
\usepackage{indentfirst,graphicx}

\begin{document}
\title{Equation-free Dynamic Renormalization of a KPZ-type Equation}
\author{David A. Kessler}
\affiliation{Dept.~of Physics, Bar-Ilan University, Ramat-Gan Israel}
\author{Ioannis G. Kevrekidis}
\affiliation{Dept.~of Chemical Engineering and Program in Applied
and Computational Mathematics, Princeton University,
Princeton, NJ  08544}
\author{Ligang Chen}
\affiliation{Dept.~of Chemical Engineering, Princeton University, Princeton,
NJ 08544}
\begin{abstract}
In the context of equation-free computation, we devise and implement a procedure
for using short-time direct simulations of a KPZ type equation to calculate the self-similar solution for its
ensemble averaged correlation function.
The method involves ``lifting" from candidate pair-correlation
functions to consistent realization ensembles, short bursts of
KPZ-type evolution, and appropriate rescaling of the resulting 
averaged pair correlation functions.
Both the self-similar 
shapes and their similarity exponents are obtained at a computational
cost significantly reduced to that required to reach saturation in
such systems.

\end{abstract}
\maketitle
\section{Introduction}

Often we are faced with the study of systems described by 
a given fine scale, microscopic dynamics, for which we would like to obtain
coarse-grained, macroscopic information.  
Such information can include stationary-states, instabilities,
and bifurcations.  
When continuum equations describing the coarse-grained dynamics
are available in closed form, traditional numerical analysis tools
can be used to obtain this type of information efficiently.
Recently, much work has been devoted to developing tools for addressing
such questions in the absence of an explicit coarse-grained description
(see \cite{PNAS,GKT,Manifesto,ShortManifesto} and references therein).
These {\em equation-free} methods offer the hope of significant 
savings in storage and run-time costs
over direct numerical simulations of the underlying microscopic
dynamics.
They also allow the study of questions inaccessible to direct
simulation, for example the characterization of unstable
stationary states.
These techniques have been applied to the coarse-grained
study of systems described by microscopic evolution rules
(kinetic Monte-Carlo, Brownian and molecular dynamics,
Lattice-Boltzmann etc., see the references in \cite{ShortManifesto}).

These tools can also be applied to problems whose solutions are
macroscopically characterized by a symmetry group, such
as translational invariance (giving rise to traveling wave
solutions). 
More recently, the techniques have been extended to problems whose
macroscopic dynamics exhibits scaling; among these are 
molecular diffusion \cite{chen}, models of
self-similar transport of random Brownian particles \cite{Zou},
core collapse in stellar systems \cite{Merritt} etc.
Such problems present challenges, both conceptual and
technical, not necessarily found in the simpler traveling-wave problems.  
For example, since scale-invariance --unlike
translational invariance-- is never an exact symmetry of the microscopic
dynamics, the scaling solutions only exist in an asymptotic limit of large
system sizes and long time-scales.
Furthermore, one must correctly identify the number of free parameters
characterizing the scaling solution.

In this context we consider the scaling behavior of an equation in the 
KPZ class \cite{kpz,review}.  
This model is a paradigm for a wide class of systems whose
correlations exhibit asymptotic scaling.  
Furthermore, as opposed to the systems studied through equation-free 
methods to date, 
the scaling behavior is not present in the
first moment of the evolving field, but rather in its correlations;
in particular, we will study the scaling solutions for the two-point
correlation function.  
We demonstrate, using the ideas outlined above, how to determine the
correlation function self-similar shape 
and the scaling exponents that characterize it.

The paper is organized as follows: in Section 2 we present the exact form
of the equation to be solved, define its correlation function and 
briefly review its known scaling properties. 
Following that, in Section 3
we discuss an iterative method for the determination of the self-similar
solution and its exponents. 
Section 4 presents a matrix-free fixed-point 
approach to the same problem, and we conclude with a discussion in Section 5.

\section{Preliminaries}
The equation we will analyze herein is a discretized form of a modified
KPZ equation (in 1+1 dimensions) for the height $h$ of an interface:
\begin{equation}
\dot h(x,t) = h''(x,t) + \frac{\lambda h'(x,t)^2}{1 + \mu h'(x,t)^2} + \eta(x,t) ,
\end{equation}
where the noise $\eta$ is $\delta$-correlated in space and time:
\begin{eqnarray}
\langle \eta \rangle &=& 0 \nonumber \\
\langle \eta(x,t) \eta(x',t') \rangle &=& 
S \delta(x-x') \delta(t-t').
\end{eqnarray}
The $\mu$ term is present to ensure that the equation does not exhibit
the finite-time singularity known to arise for sufficiently large $\lambda$
in the unmodified $\mu=0$
equation \cite{yuhai}. 
Dimensional analysis shows that $\mu$ does
not change the scaling behavior of the system.

In discretized form, our equation reads
\begin{equation}
h^{t+\Delta t}_i = h_i^t + \Delta t\left[h^t_{i+1} - 2h^t_i + h^t_{i-1} + \frac{\lambda(h^t_{i+1} - h^t_{i-1})^2}
{4 + \mu(h^t_{i+1} - h^t_{i-1})^2} + \eta^t_i\right]
\end{equation}
with  $\langle \eta^t_i \rangle = 0$, $\langle \eta^t_i \eta^{t'}_j \rangle =
S \delta_{t,t'}\delta_{i,j}/\Delta t$. 
We work in a periodic box of length $L$.  
In the following, we take $\mu=1$, $S=1/12$, $\Delta t=0.05$. 
It is customary to start simulations ($t=0$) from a flat interface $h^0_i=0$.
It is convenient to define the
average height $\bar{h}^t \equiv \frac{1}{L}\sum_{i=0}^{L-1} h^t_i$ and the
reduced height $\hat h^t_i \equiv h^t_i - \bar{h}^t$.  
Since a flat interface
is stable, $\langle \hat{h_i} \rangle =0$, and the basic object of interest
is the two-point correlation function: 
\begin{equation}
G^t(d) = \frac{1}{L}\sum_{i=0}^{L-1} \langle \hat{h}^t_i \hat{h}^t_{i+d} \rangle
\end{equation}
or, equivalently, its (discrete) Fourier transform:
\begin{equation}
G^t_K = \frac{2}{L}\sum_{d=0}^{L/2} e^{2\pi i Kd/L} G^t(d)
\end{equation}
where $K=kL/(2\pi) = 1,\ldots,L/2$ and  $k$ is the momentum.

Let us briefly review the basic known scaling properties of $G^t_K$
\cite{review}.  
For intermediately large wavenumber $K$, $K^*(t) \ll K \ll L/2$, $G^t_K$ falls like
a power-law for increasing $K$, $G^t_K \sim C/K^{2\alpha + 1}$, where $C$ is
independent of $t$.  
At small $K \ll K^*(t)$, the power-law growth is cut off, and 
$G$ approaches a finite limiting value as $K \to 0$ with zero
derivative.
The crossover length scale, $1/K^*(t)$, grows with time as 
$t^{1/z}$, until the saturation time when $1/K^*(t) \sim L$. 
This saturation time thus grows with the system size as $L^z$. 
At small $K \ll K^*(t)$, $G$ saturates at a value
which grows in time like $t^{(2\alpha+1)/z}$, again until the 
saturation time.  
The values of the exponents $\alpha$ and $z$ in this one-dimensional
system are known  analytically , $\alpha=1/2$, $z=3/2$. 
Putting this all together, $G$ has the scaling form
\begin{equation}
G^t_K = t^{(2\alpha+1)/z} f(Kt^{1/z}).
\end{equation}
where the function $f(x)$ approaches a constant for small arguments and
decays as $1/x^{2\alpha+1}$ for large $x$.

\section{Solution by Direct Iteration.}
Our goal is to find a self-consistent $G_K^t$ with the correct 
scaling properties, i.e. find the function $f(.)$ in Eq. 6 above.
Roughly speaking, if we knew the exact $G_K^t$, we could use this knowledge
to generate an ensemble of initial $h$ fields at some time $t_0$, 
conditioned on this $G$.  
We could then evolve each of these initial 
$h$'s forward
in time to some $t_f$, and calculate $G$ at this later time.  
The new $G$
should then simply be a rescaled version of the original $G$.  
This scaling
condition is what we use to determine $G$.

There are a number of caveats that need to be expounded at this point.  
Some
of these are technical in nature, involving the details of the algorithm
used to actually find $G$.  
Two, however, involve matters of principle.  
The first caveat is that the full statistical solution is not uniquely 
determined solely by the two-point function $G$.  
In principle, a complete
solution implies knowledge of all higher-order correlations as well.  
However, {\it we posit} that the dynamics is such that the 
higher-order correlators relax much more quickly
than the two-point function itself.  
If this is true, there will be a
short transient during which time the system will 
reconstruct all the higher-order correlations (will ``heal")
while $G$ itself does not change much.  
In effect, this is an assumption of separation of time scales
between different order correlators.
Thus, to determine
the solution, we compare not $G$ at times $t_0$ and $t_f$, but at times
$t_I$ and $t_f$, where $t_I$ is an intermediate time chosen so that the
higher-order correlators have had time to become slaved to $G$ (recover 
their correct values conditioned on the given $G$).

The second caveat is that $G$ exhibits the desired scaling properties 
{\it only asymptotically}.
At the smallest scales, the underlying lattice ruins scale-invariance.
More serious, however, is the fact that for short times and small scales,
the scaling properties of $G$ are those of the Edwards-Wilkinson, i.e., the
$\lambda=0$ model \cite{EW}.  
Only for sufficiently large times is the interface
sufficiently rough for the nonlinearity to determine the scaling. 
Starting simulations from a flat interface and evolving for time $t_0$
brings the interface to some scale of roughness parametrized by $t_0$.
Our procedure will converge on the particular member of the scaling 
family that has this characteristic roughness scale; needless to say,
all members of the scaling family (at large enough scales !) can be
directly reconstructed.
As we will see in more detail below, unless the initial time $t_0$
(alternatively, the initial roughness scale) is large enough, our
results will be contaminated by the short-time, small scale Edwards-Wilkinson
behavior.
Only if our working scale is large enough for the asymptotically
self-similar behavior to have set in, does our process make sense.
In practice this means that we must test the working roughness scale
(alternatively, the working $t_0$) to confirm that both the self-similar
shape and the exponents have converged.
As we increase $t_0$, we will of course have to increase $L$ accordingly, so
that we continue to capture the full self-similar structure of $G$.

We now move on to the technical details of the calculation.  
The primary issue
to address is how to solve the fixed-point equation embodying the 
scale invariance condition.
We will do this in two different ways.  
The first is by successive substitution.  
Here we just perform repeated cycles of forward integration,
followed by rescaling to the original time (roughness).  
The second is via a fixed point Newton-type procedure, which we describe in
the next section. 

Direct iteration proceeds as follows.  
We start by integrating the
system forward from a flat interface at $t=0$ to some $t_0$.  
We then calculate $G^0_K$.  
It is
straightforward to generate configurations $h_i$ conditioned on this $G^0_K$
(to ``lift" from $G$ to $h$).
All we have to do is to remember that $G_K$ is the expected value of
$|h_K|^2$. 
Thus, we generate a Gaussian random number with zero mean and
variance $G^0_K$, and multiply by a random phase to obtain an $h_K$.  
An inverse Fourier
transform gives us our desired $h_i$.  
We generate some number, (typically
32,000) of such initial configurations and integrate each forward in time to
$t_f$, which we take to be $t_f=2t_0$, and measure $G^f_K$.  
We also measure $G^I_K$ at an intermediate time $t_I=3t_0/2$. 
From these measurements we construct the {\it functions} 
$G^0(K), G^I(k)$ and $G^f(k)$ (see below for details).
We now need to rescale
$G^f(k)$ back to the original roughness scale (or time $t_0$).  
We do this is two steps.
We first determine, for each measuring time,
a typical small $k$ scale, $k_{1/2}$ by the condition
$G({k_{1/2}}) = G(0)/2$. 
We next rescale  wavenumbers at times
$t_I$ and $t_f$ by the factors $f_k^{I,f}=k_{1/2}^0/k_{1/2}^{I,f}$,
respectively; this ``aligns" their rescaled large scale behavior.
We next rescale $G^I$ and $G^f$ by factors $f_G^{I,f}=
G^{I,f}({k_B/f_k^{I,f}})/G^0({k_B})$ (where $k_B=\sqrt{L/2}$ is
the geometric mean of the smallest and largest wavenumbers)
so that their large wavenumber behavior coincides with that of $G^0$.
For the self-similar shape, the function $G^f(k/f_k^f)/f_G^f$
should reproduce our original function $G^0(K)$.  
In practice, it gives us
a new starting point for another round of iteration; in our 
problem the self-similar solution is stable,
and this procedure rapidly converges, so that in fact 
the differences are only due to fluctuations.
If the problem was truly (as opposed to asymptotically) self-similar,
upon convergence to the self-similar solutions,
the scaling factors $f_K^{I,f}$ and $f_G^{I,f}$ could be used to estimate 
the effective scaling exponents $\alpha$ and $z$:
\begin{eqnarray}
z&=&\ln(t_f/t_I)/\ln(K_{1/2}^I/K_{1/2}^f)=\ln(t_f/t_I)/\ln(f_K^f/f_K^I)\nonumber\\
\alpha &=& \frac{1}{2}\left(\frac{\ln(f_G^f/f_G^I)}{\ln(f_K^f/f_K^I
)}-1\right)
\end{eqnarray}
Because the self-similarity is only asymptotic, three successive times are
in fact needed to estimate the exponents.
This procedure constitutes the equation-free implementation of dynamic
renormalization (see e.g. the classical references \cite{McLaughlin,LemesurierA,LemesurierB},
as well as our template-based approach discussed in \cite{Aronson,Rowley2}).

The last point we need to cover is how we convert our measured $G_K$ into
a function $G(k)$.  
We use a relatively low-dimensional description of the function G(k);
in particular, we fit a cubic spline to $log(G_K)$ throughout the whole
range, with knots approximately equally spaced in $log(K)$; this provides
an adequate description with $O(10)$ degrees of freedom.

\begin{figure}
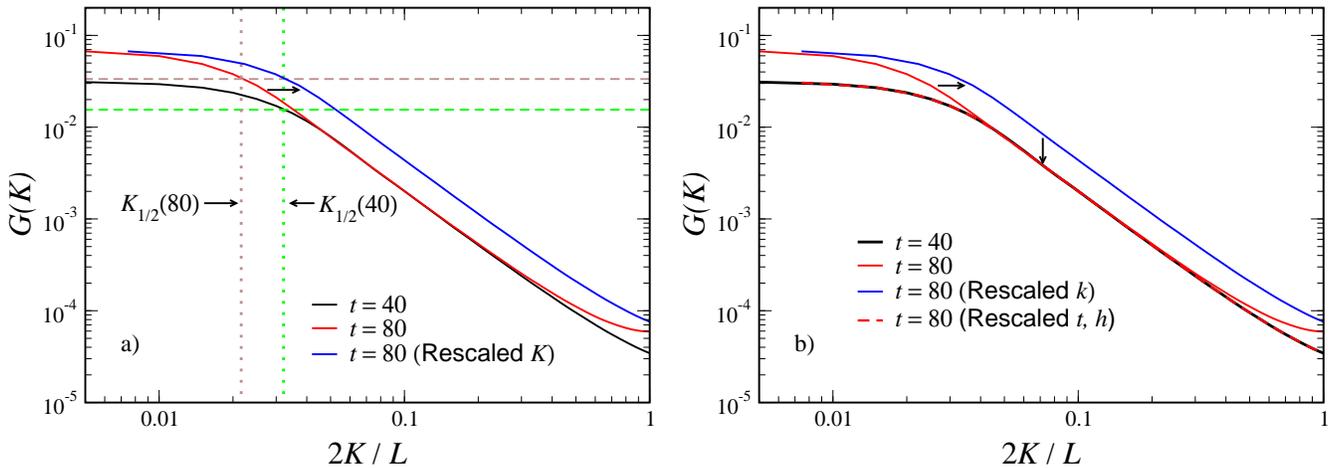

\includegraphics[width=3.4in]{t40_80d.eps}\quad
\includegraphics[width=3.4in]{t40_80e.eps}
\caption{The iteration cycle for $\lambda=5$, $L=400$, $t_0=40$.  a) $G(K,t)$ for  $t=40$ and $t=80$, with their respective $K_{1/2}$'s noted, 
 together with $G(K,80)$ after rescaling of $K$ to align its $K_{1/2}$ with $G(K,40)$. b)$G(K,40)$ and $G(K,80)$, together with $G(K,80)$ after rescaling of $K$ and after rescaling of  both $k$ and $h$ to collapse it onto $G(K,40)$.}
\label{cycle}
\end{figure}

An example of the results of this procedure are shown in Fig. \ref{cycle}, where
one cycle of the iteration is shown.  
We see that indeed $G(k)$ is
recovered to quite good accuracy by our integration and rescaling,
except for the very largest $k$'s, where the zero-slope condition imposed by
the discreteness of the lattice is evident in the original $G$, but
not in its rescaled version.  

\begin{figure}
\includegraphics[width=4in]{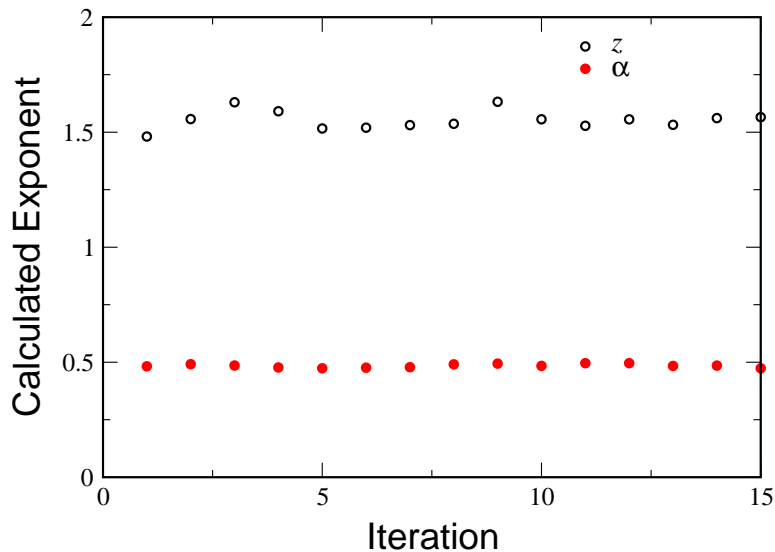}
\caption{Calculated values of the exponents $z$, $\alpha$ vs. iteration for $\lambda=5$, $L=400$, $t_0=40$.}
\label{l400t40}
\end{figure}

\begin{figure}
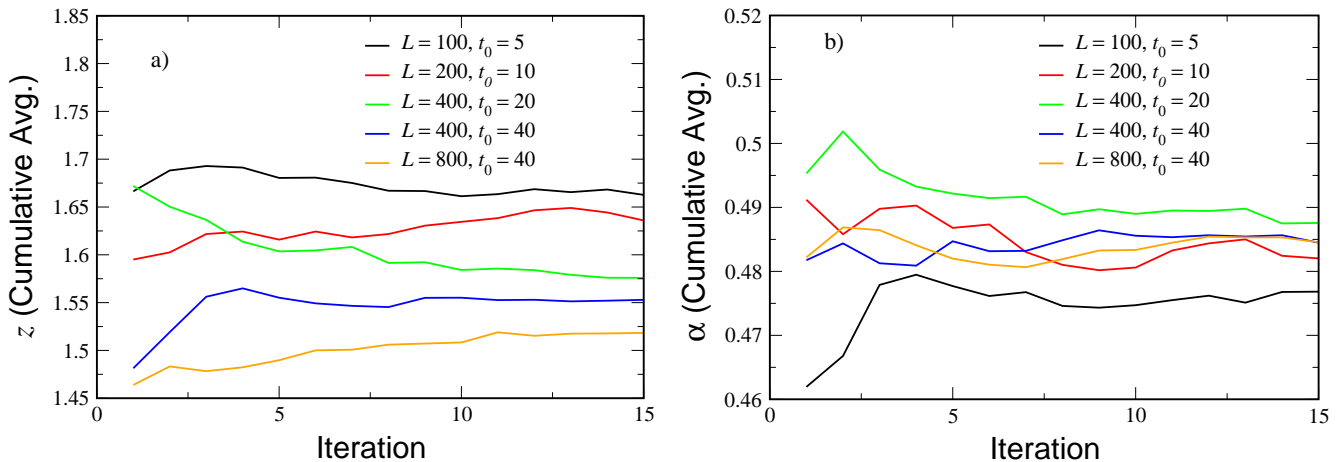

\includegraphics[width=3.4in]{z_cum.eps}\quad\includegraphics[width=3.4in]{alf_cum.eps}
\caption{Cumulative average of the estimated exponents (a) $z$; and (b)$\alpha$ as a function of iteration number
for $\lambda=5$ and various pairs of $t_0$, $L$.}
\label{cum}
\end{figure}

\begin{figure}
\quad\includegraphics[width=4in]{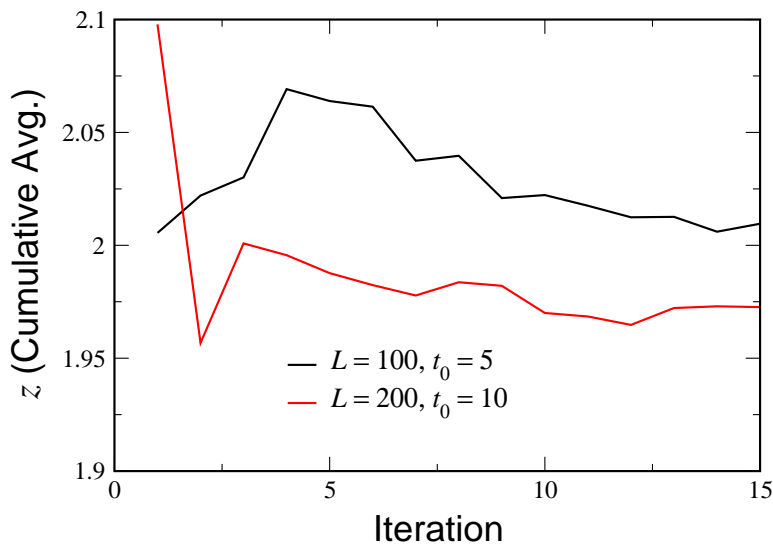}
\caption{Cumulative average of the estimated exponent $z$ as a function of iteration number
for the Edwards-Wilkinson model ($\lambda=0$) and two sets of $t_0$, $L$.}
\label{ew}
\end{figure}

We present in Fig. \ref{l400t40} a graph of the estimated
exponents as a function of iteration number, for some particular
value of $t_0$ and $L$.  
We see that the iteration 
is quite stable, albeit with sizable statistical fluctuations.  
In Fig. \ref{cum}(a-b), we present the cumulative average over iterations
of the estimated exponents, as a function of iteration number, for various
sets of $t_0$ and $L$.
What is remarkable is that the calculation essentially converges 
immediately, to the precision we can measure.  Our procedure allows us work at an intermediate
scale where we can see simultaneously the rapid saturation of the correlation function at short length scales and
the growth at the coarse scales, tracking the shift of the crossover between these two regimes while this evolution is
still relatively fast; approaching the ultimate shape while
    the interface evolves to coarser scales would take a significantly longer computational time (the coarser the scale, the
longer the time).
    It is this ``shape evolution at constant scale" that underpins
    the computational savings of the method. 
We see that increasing $t_0$ (alternatively, the working roughness scale)
brings the measured value of $z$ down much closer to its
asymptotic value of $3/2$.  
Since the value of $z$ for the
Edwards-Wilkinson
model is $2$, we interpret this as indicating the contamination of our
measurement by the crossover from Edwards-Wilkinson behavior at short
scales.  
In Fig. \ref{ew}, we present for comparison the data for the
Edwards-Wilkinson model.  
Here we immediately obtain values very close
to the expected $z=2$, $\alpha=1/2$, as the only violation to scaling
comes from the very short scale lattice effects.

\section{Newton-GMRES Fixed Point Solution}
A direct iteration procedure is, of course, not guaranteed to 
converge; the self-similar solution may not be stable for the 
system and parameter values of interest.
Even if it is stable, one needs to start in its basin of attraction,
and the rate of approach will asymptotically depend on the local
linearization characteristics of our fixed point formulation.
It is therefore desirable to develop approaches that do not rely
on the stability of the fixed point, and the Newton method is the
obvious choice.
Lack of an explicit macroscopic equation means that the Jacobian
involved in Newton iterations is not explicitly available.
In principle one could estimate the Jacobian using finite differences;
yet, especially for large scale and noisy problems, such an estimation
will be both prone to error and very costly.
Krylov-subspace methods, such as GMRES, have been devised for the iterative
solution of linear equations; they are based on the evaluation of matrix-vector
products  of the system Jacobian with a sequence of algorithmically
determined vectors.
When the Jacobian is not explicitly available, matrix vector products 
can be estimated in a {\it matrix-free} fashion from nearby function
evaluations - this is the basis for matrix-free Newton-Krylov-GMRES
algorithms \cite{Kelley}.
Such algorithms are naturally suited to the fixed point problems arising
in our equation-free renormalization scheme - lifting from nearby 
two-point correlation functions, evolving through the KPZ dynamics,
and rescaling the resulting averaged two-point correlations results in an estimate
of the action of the Jacobian of our fixed point problem.
This type of iteration is often particularly well suited for problems with
a separation of time scales \cite{Liang}.

We used Newton-Krylov GMRES iteration to find the shape of the fixed point
of the KPZ renormalization problem; the operating parameters were $L=400,
\lambda=5$, and a short burst of time simulation, $\Delta t=5$, was used 
to calculate the evolved shape.  
Cubic spline interpolation was again used in the construction of the
function $G(k)$. 
All the data were averaged over $96000$ replica initializations
consistent with this $G(k)$.
The initial guess was obtained by starting from a flat interface
and evolving for time of $t_0=5$.
As shown in Fig. 4, it took roughly three iterations for the plotted
norm of the residual to decrease by one order of magnititude.
The resulting fixed point is visually indistinguishable from the one arrived
at by direct substitution, and the same scaling exponents, within the
fluctuation bounds, are recovered.
The Newton-based process is not computationally efficient in problems such
as this, where the self-similar solution is strongly attracting; it
would become advantageous, however, in cases where the self-similar solution 
is very slowly attracting, or simply unstable, when direct iteration will
not converge at all.

\begin{figure}[htbp]
\centering
\leavevmode
\includegraphics[height = 2.5in]{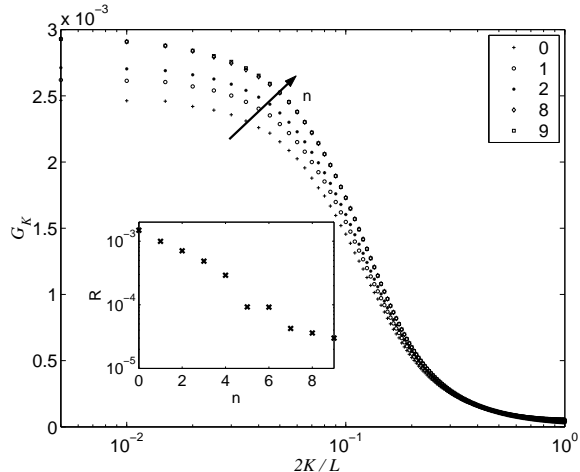}
\caption{GMRES iterations to find the fixed shape ($L=400, \lambda=5,
\Delta=5, N=96000$)  The inset shows the residual $R$ as a function of iteration number.}
\label{fig3}
\end{figure}

\section{Discussion and Conclusions}
We presented a computational approach to finding self-similar 
solutions {\it for the statistics} of a KPZ-type stochastic evolution
equation.
This was accomplished without an explicit, closed form of an equation
governing the dynamics of these statistics (in particular, the two-point
correlation function); we {\it estimated} this unavailable equation on 
demand
using short bursts of appropriately initialized direct simulations.
The approach was successful in reproducing the known scaling behavior 
of the model. 
One of the important features of the computation was the detection of
the signature of Edwards-Wilkinson scaling in the data when the 
working scales were not chosen large enough (due to asymptotic self-similarity).
In our current implementation we used two {\it local} conditions 
(``templates", \cite{Rowley1,Rowley2}) to implement
the two rescalings, those of the stretching of the wavenumber (or reshrinking of the length scale) and
the scaling down of $G_K$ (equivalently the field variable amplitude $h_K$). 
It would be preferable to
employ non-local conditions, making the computation more robust to
local fluctuations through averaging.

Perhaps the most important assumption of the equation-free approach is that
an equation exists {\it and closes} at a chosen level of description;
here we assumed that the appropriate level was the two-point correlation
function, and that higher order correlations either do not affect this evolution
or become quickly slaved to it. 
It is known \cite{edwards-schwartz} that in one spatial dimension the exponents are insensitive
to variations in the third and higher order correlators; this appears not to
be true in higher dimensions.
In our computational experiments we found that third order correlations did
not become quickly slaved to two-point correlations (over times comparable 
to the evolution of the two-point correlation itself); since we worked in 
1+1 dimension, this is consistent with the above observation.
In higher dimensions, it is not clear that an equation does indeed close in
terms of only the two-point correlation function; testing this hypothesis
would be an important first task to pursue with our approach.
Computational approaches to initializing ``fast" variables consistently with
slow ones (alternatively, on a manifold parametrized by the slow ones) have
long been known in computational chemistry (\cite{Ciccotti,Valleau}), and 
their use in an equation-free context is discussed, for example, in \cite{GearK}.

In this paper we used a KPZ-type SDE as our ``inner", fine scale solver.
The procedure is identical if the SDE solver is substituted by, for example, a
kinetic deposition model; the computation of coarse self-similar 
solutions for such models is underway.
For some of these, such as ballistic aggregation, we do not anticipate any difficulties;
for other, more highly constrained models, e.g. the restricted SOS model \cite{review}
the lifting operation should be nontrivial.
In the case of stable self-similar solutions 
additional equation-free techniques, such as coarse projective integration \cite{GK1,GK2}
can be used to accelerate the computation of self-similar dynamics.

An important test of the approach will be its ability to compute and characterize
the unstable fixed point that is known to exist in $d \ge 3$ dimensions; this is
a case where the Newton-GMRES procedure would be crucial.

{\bf Acknowledgements}  This work was partially supported by NSF and DOE (USA) and by the Israel Science
Foundation.

\end{document}